\documentstyle[12pt]{article}
\begin{document}
\title{ Complete Quantum Communication with Security} \author{Arindam Mitra
\\Lakurdhi, Tikarhat Road, Burdwan, 713102,\\
 West Bengal,  India.}

\maketitle
\begin{abstract}
The long-standing problem of quantum information processing 
is to remove the classical channel from quantum communication.
Introducing a new information processing technique, 
it is discussed that both insecure and secure quantum communications
are possible without the requirement of
classical channel. 
 
\end{abstract} 

\section*{}            
Around 1970, Wiesner first realized [1]  quantum state 
can be used to ensure security of private data. Security
reason apart, the idea of using quantum state for information processing
was new, but because of delayed publication it then silently laid the foundation of modern quantum 
information. However,  quantum information
received widespread attention
 after the discovery
of quantum key distribution (QKD) [2, 3],  quantum computation [3],
teleportation [4]. Except the {\em local} quantum computation, all are basically
quantum communication protocols. To execute these protocols
a supplementary classical channel is required.
In other words, in quantum communication,
classical channel is needed 
to transmit message. One may inquire  why quantum mechanics cannot provide
a quantum channel to transmit  message.\\

In their famous paper on teleportation [4],
Bennett {\em et al} have apparently given some answer to this query.
They argued  that  unknown quantum state cannot be teleported without 
classical channel. If it is possible it will prove the existence of
superluminal signal. The intended meaning and reasoning
can be easily understood but the  problem is,
they have implicitly assumed that
classical channel is  the only
legitimate tool for sending logical 0 and 1 to transmit message.
And now the assumption  has almost become irrefutable theorem. Here we shall see that logical 0 and
1 can also be coded and decoded in  quantum fashion.
It means, a qubit  can represent logical 0 and  1.
 It implies that quantum channel can alone transmit message,
 and in quantum computation,
 controlled gates can be controlled quantum mechanically
 by receiving human instructions. \\

Before constructing our quantum channel
let us  classify the classical channels.
This might help explore the reason behind their assumption. \\
1. Deterministic classical channel:  Here every  
bit can be deterministically recovered. Example: existing
classical channel in absence of noise.\\
2. Probabilistic classical channel: Here every bit can be
statistically recovered. Example: two induced noise levels
greater than  the environmental noise can represent the two bit values (
this channel [5] can provide {\em computational} security).\\

Let us construct  channels using quantum states.\\
1. Channel A: Two orthogonal quantum states
represent the two bit values.\\
2. Channel B: Two nonorthogonal quantum states
represent the two bit values. \\
3. Channel C:  Two ensembles of quantum states
represent the two bit values (in ensemble interpretation,
however, individual
state has no  place  as
perceived by Einstein[6]).\\

The channel A is a deterministic channel and classical message
can be sent through it. The channel B is a probabilistic quantum channel
and  message cannot be sent through it [7].
The channel C is another kind of probabilistic channel but  message
can be sent through it.  The channel A is quantum counterpart
of deterministic classical channel and the channel C is
quantum counterpart of probabilistic classical channel. These two channels
cannot provide quantum encryption  where no-cloning principle [8] provides 
 security. For Channel  A, no-cloning principle breaks down and for 
channel C it is not applicable. As a whole,
these two channels do not carry any signature of quantum information
processing, and therefore, they may not be called as truly quantum channel
(like  quantum states do not necessarily perform quantum computation).
From the above discussion
it seems that truly  quantum channel does not exist
to transmit plain or secure  message. \\   

We shall see that such channel do exist. To construct such
quantum channel we need two sufficiently distinguishable  sequences of quantum states to
represent the two bit values. The most simple sequences are the
sequences of two nonorthogonal states, say $0^{\circ}$ and $45^{\circ}$
single photon polarized states. Suppose in the sequence, representing bit 0,
$0^{\circ}$ and  $45^{\circ}$ photons are at the even and odd positions
respectively, and in the sequence representing bit 1,
 the   $0^{\circ}$ and $45^{\circ}$ photons are at odd and even positions
 respectively. Let us also assume that receiver share the information of these
two sequences and only a single bit - a  sequence - is sent.
The question is, how receiver will
definitely recover the bit value by  measurements.
Suppose receiver sets the analyzer  at $0^{\circ}$ for every
even event. If  he/she gets "yes" results at every even position then the bit is
definitely 0. If he does not get so, the  bit is definitely  1.
This is a new type of probabilistic quantum channel 
but not secure quantum channel, and message can be sent through it. 
Message can be seen as a sequence of two operating sequences
of quantum states.
For quantum security, legitimate users  have to generate a arbitrarily long
sequence
of  randomly operating two sequences of random states
by
secret sharing of information
of the two operating sequences, although the randomness of states of the
two operating sequences is not a stringent criteria.
 
\paragraph{}As  for example, let us take the following two operating 
sequences of two different pairs
of random quantum states.\\ 
\noindent
$S_{0}
=\left\{\vert A \rangle_{1},\,\vert A \rangle_{2},\,\vert B \rangle_{3},
\,\vert B \rangle_{4},\,\vert A \rangle_{5},\,\vert B \rangle_{6},\,
\vert B \rangle_{7},\,\vert A \rangle_{8},......\vert A \rangle_{n}\right\}$;\\
\noindent $S_{1}=\left\{\vert C \rangle_{1},\,\vert D \rangle_{2},\,\vert C \rangle_{3},
\,\vert D \rangle_{4},\,\vert C \rangle_{5},\,
\vert D \rangle_{6},\,\vert C \rangle_{7},\,\vert D \rangle_{8},.....
\vert D \rangle _{n}\right\}$,\\
 where  $S_{0}$  and  $S_{1}$  stand  for  bit  0  and   1
respectively and $n$ is moderately large number.
Information of these  two  sequences  $S_{0}$  and $S_{1}$ are shared
between sender Alice, and receiver Bob.  Key,  the
sequence  of  random sequences of random quantum states, is,  
\noindent $K_{N} =\left\{S_{0}, \,S_{1},
\, S_{1}, \, S_{0}, \, S_{1}, \, S_{0}, \,  S_{1},
\,  S_{0}, \, S_{0}, \, S_{1},.......\right\}$ , where $N$
is the number of  bits in the key. Obviously, N is greater than 2n since
2n {\em bits} (standard meaning) are shared.\\

First, we shall present a  QKD protocol using superposition states
for the preparation of the two operating sequences. The reason is,
if we can construct the channel by superposition states and if it gives quantum
security then we do not have any
ambiguity about the quantumness of the channel.
This particular QKD protocol can be modified to accomplish more
sophisticated cryptographic tasks such as
key splitting [9,10] and quantum bit commitment [11].

\paragraph*{}
Firstly, we describe the preparation procedure of the two  sequences. 
Suppose, in a secret place, Alice and Bob have  $2n$  horizontally  polarized  $(\vert{\leftrightarrow}\rangle)$
incoherent photons. To prepare a sequence they use n photons.
To  prepare  $S_{0}$, they split the wave function of each photon
 with a symmetric (50:50) beam splitter. After splitting they
do  one of the two things in one of the path, called  {\bf s} : toss a coin, and
if the result is
"tail"  they   do
nothing  $(\vert{\leftrightarrow}\rangle_{\bf s} \,\longrightarrow \,
\vert{\leftrightarrow}\rangle_{\bf s})$ and if
"head", unitarily rotates  the  polarization  by
$90^{  \circ}  (\vert{\leftrightarrow}\rangle_{\bf s} \longrightarrow
\,  \vert{\updownarrow}\rangle_{\bf s})$.
 In the other  path,  called
{\bf  r},  they  do  nothing $(\vert{\leftrightarrow}\rangle_{\bf r}
\longrightarrow\vert{\leftrightarrow}\rangle_{\bf r})$. They repeat
this procedure for n photons.  The   states
are        :\begin{eqnarray}       \vert A \rangle_{i}       =
1/\sqrt{2}(\vert{\leftrightarrow}\rangle_{\bf r}                    +
\vert{\leftrightarrow}\rangle_{\bf s})          \nonumber          \\
\vert B \rangle_{i}                                          =
1/\sqrt{2}(\vert{\leftrightarrow}\rangle_{\bf r}                    +
\vert{\updownarrow}\rangle_{\bf s})\nonumber \end{eqnarray}\\
 To  prepare $S_{1}$, similarly  after splitting a wave function
 they do one of the two  things
in the path {\bf\ s} : toss a coin; if "heads", unitarily rotates
by         $45^{\circ}         (\vert{\leftrightarrow}\rangle_{\bf s}
\,\longrightarrow                                              \,
\vert{\nearrow\!\!\!\!\!\!\swarrow}\rangle_{\bf s})$  and  if "tail",
unitarily        rotates         by         $135^         {\circ}
(\vert{\leftrightarrow}\rangle_{\bf s}
\,\longrightarrow\,\vert{\nwarrow
\!\!\!\!\!\!\searrow}\rangle_{s})$.
 Similarly in the other path {\bf r}, they do nothing. They repeat
 this procedure for n photons. The states
are        :\begin{eqnarray}       \vert C \rangle_{i}       =
1/\sqrt{2}(\vert{\leftrightarrow}\rangle_{r}                    +
\vert{\nearrow\!\!\!\!\!\!\swarrow}\rangle_{s})    \nonumber   \\
\vert D \rangle_{i}                                          =
1/\sqrt{2}(\vert{\leftrightarrow}\rangle_{\bf r}                    +
\vert{\nwarrow\!\!\!\!\!\!\searrow}\rangle_{\bf s})         \nonumber
\end{eqnarray} By the above operation,
they essentially
prepare the two operating sequences of superposition states.
These superposition states are mutually nonorthogonal states
and they can be represented by the following
base                   states:                   \begin{eqnarray}
\vert{\leftrightarrow}\rangle_{\bf r},\,
\vert{\updownarrow}\rangle_{\bf r}                                ,\,
\vert{\leftrightarrow}\rangle_{\bf s}
,\,\vert{\updownarrow}\rangle_{\bf s}  \nonumber  \end{eqnarray}   In
this   basis,  the  density  matrix  of  the  two  sequences  is,
\begin{eqnarray} \rho =1/4\left(\begin{array}{clcr}2 & 0 & 0 &  0
\\0  &  0  &  0  &  0\\0  &  0  &  1  &  0  \\0  &  0  &  0  &  1
\end{array}\right)\nonumber\,\,   \end{eqnarray} 

\paragraph*{}
 Now they are separated. Alice will transmit the
two operating sequences at random. For clarity, let us think that
 Alice transmits a single bit, either $S_{0}$ or $S_{1}$. Bob's 
task is to recover the bit.
 As density matrices are same, one may think, how Bob will recover it.
Bob can  independently   recover the bit value in  different  ways
since  he  knows the preparation code  of the both types of possible bits,
although their density matrices are same
(note that, $\rho_{0} =\rho_{1}$ for earlier example).
Whatever be the identification  processes, Bob's objective is to recover
the bit value from the shared information. Basically there are
two types of measurement tricks: \\
 1. Sequence of measurements is predetermined according
to the preparation codes.\\
 2. Sequence of measurements is not predetermined
according to the preparation codes.  \\

In  this protocol,
we  shall use conclusive {\em which path} (WP) information
to recover the bit value by using the second method.
But mere $WP$
information  is  not  enough  to identify the states and bit values.
Bob needs {\em which-path of which-state} ($WPWS$)  information
to identify the individual states and bit values. But $WPWS$  information
is also not enough , he  needs {\em which-path of which-state of which
density matrix} (WPWSWD) information to identify the states and
the bit values (here {\em which density} matrix means
how it was prepared).
Next we shall see how to get
WPWS information.

\paragraph*{}
Suppose there are two sets
of  dual  analyzers  ($DA$)  on  the  end of the two  resulting  paths.  Suppose
the
orientations  of  $DA$  are : i)  $DA_{0}$ =
 $\left\{0^{\circ}_{r}    :   0^{\circ}_{s}\right\}$   ii)   $DA_{1}$   =
$\left\{0^{\circ}_{r} : 45^{\circ}_{s}\right\}$.
 The measurements produce three types ($\alpha, \beta, \gamma$) of results: 
$\alpha$
=     $(\surd_{\bf r}     :     \times_{\bf s})$,\,$\beta$=     $(\times_{r}:
\surd_{\bf s})$,\,$\gamma$ = $(\times_{\bf r}:  \times_{\bf s})$  
where  $"\surd"$
and  $"\times"$  stand for " yes" and " no" results respectively.
The probabilities of these three kinds of  results  corresponding to
  the  four
different  superposition  states  are  given  in  table 1 and 2
considering the statistical weight of the states and orientations
of DA.
The results $\alpha$ and $\beta$
provide  {\em which-path} ($WP$) information and the result $\gamma$
gives {\em no-path} ($NP$)information.
The result  $\alpha$  does  not give any $WPWS$ information for any of the
above two settings of $DA$. The $NP$ information, corresponding to
the  result $\gamma$, is  
always  inconclusive for any of the two settings of the $DA$.
The only result $\beta$
provides conclusive $WPWS$ information for proper choice
of above two
settings of $DA$. The $WPWS$ information  conclusively determines 
the state $\vert A \rangle_{i}$ for $DA_{0}$ and the state
$\vert C \rangle_{i}$ for $DA_{1}$. But we need {\em WPWSWD} information
to recover the bit value.

\paragraph*{}  As  Bob  does  not  know the bit value in advance,  he
 uses two sets of $DA$ at random according to the second trick. The measurements  yield  two
sets  of  random results. Firstly,  Bob discards inconclusive
results ($\alpha$ and $\gamma$) from  both  the  sets. For the time being, let us assume that the
states $\vert A \rangle_{i}$ are at even positions
and  the states $\vert C \rangle_{i}$ are at odd positions in the sequence $S_{0}$
and $S_{1}$ respectively.
If the bit is 0, the reduced sequences of results will like this 
(see the 2nd col. of the tables),\\
$P^{n/8}$=$\left\{...,\beta_{4},.....,\,\,
\beta_{10},.......,\beta_{18},....,\,\,\beta_{32}........\right\}$ for $DA_{0}$, \\
If the bit is 1 then it will look like: \\
$Q^{n/8}$ 
= $\left\{...,\beta_{3},....,\,\,\beta_{11},.....\,\,\beta_{21},.....,
\,\,\beta_{27},........\right\}$ for $DA_{1}$\\ 
where n/8 is the reduced 
length of the sequences( 75\% results are discarded). 
It means Bob will only check whether $\beta$ is missing in odd or even possitions
of his two reduced sequences of results to recover the bit value. 
But the state $ \vert A\rangle_{i}$ and 
$\vert C \rangle_{i}$ will not have  regular distributions;
they are randomly distributed. Yet the technique
of recovery of bit value is almost same. 
 Bob first
discards the states corresponding to the discarded  events  from
both of the shared  sequences.  
Now  with  these two reduced sequences
of results ($P^{n/8}$ and $Q^{n/8}$) 
and two reduced sequences of states ($S^{n/8}_{0}$ and $S^{n/8}_{1}$), 
he performs  four  
correlation tests to identify the bit. Out of these tests, only one 
of the reduced sequences of results  will be totally correlated  
(assuming noise is not present) with one of  the  reduced sequences 
of states. Recovery of the bit value means recovery of WPWSWD information.\\  

\noindent 
{\bf Security}: First, we want security directly from no-cloning principle.
It means we need bit by bit by (sequence by sequence) security
as this  is a repetitive code.
 In bit by bit security,  Alice will not send the
 next sequence, until she is confirmed that Bob's received sequence was
 uncorrupted.  Therefore, after transmission of each bit, the
feedback communication  is necessary.
This feedback communication
should be authentic communication.
Like existing QKD schemes, if we want to use classical
authentication [12] we need some additional secret data.
It implies that  a single created data can be made secure by
using more secure  data! This is totally unrealistic.  We need new 
kind of authentication technique. \\

\noindent
{\bf Quantum authentication (QA)}: For authentication,   following
feedback technique can be taken.
If  the received sequence is uncorrupted, Bob
can  send back the same sequence or the other  to Alice.
If  Alice also found it uncorrupted she will send the next sequence to Bob.
Otherwise Alice  will  stop  transmission
and reject the two operating sequences.
There is no room for peaceful co-existence with  eavesdropper.
In contrast, in existing QKD, one can compromise with eavesdropper.\\

\noindent
{\bf Silent features}:\\
1. Security and authentication are simultaneously achieved by the protocol 
itself. Nowhere it has been revealed. \\
2. For QKD, classical authentication and classical channel cannot be
used (at least classical channel
can be used for QA in one of our entanglement-based scheme [13])\\
 3. If the users do not share information of the two operating sequences,
still key/message can be recovered
at the cost of security. As if receiver impersonates
as eavesdropper to clone the key.\\

\paragraph{Key splitting :} Key splitting is one of the important task of  cryptography. The purpose is to 
distribute  a key securely to two (or many) receivers to make them mutually
dependent on each other. 
The above two-party protocol can be  
extended to perform this task. \\

Suppose  two receivers, Bob and Sonu,
in the two resulting paths, where Bob is on the path {\bf r}  and  Sonu  on
the path  {\bf  s} and both of them share information of 
the {\em same} two operating sequences  with Alice.  Notice  that, only {\bf s} 
is the bit-carrying path.
So Sonu independently can identify the bit values. That is,
she can have WPWSWD information using single analyzer with proper
orientation [see discussion in ref. 11].
But  Bob  cannot. Bob always gets the same truncated state
$\vert{\leftrightarrow}\rangle_{\bf r}$ which never carries any bit values.
To  give equal opportunity
to Bob, Alice can  make the path {\bf r} as bit-carrying path.\\

\noindent
Then the new superpostion states are:
\begin{eqnarray} \vert A \rangle_{i} = 
1/\sqrt{2}(\vert{\leftrightarrow}\rangle_{\bf s}                   +
\vert{\leftrightarrow}\rangle_{\bf r})  \nonumber \\
\vert B \rangle_{i} = 1/\sqrt{2}(\vert{\leftrightarrow}  \rangle_{\bf s}
  +
\vert{\updownarrow}\rangle_{\bf r}) \nonumber \\ 
\vert C \rangle_{i}       =
1/\sqrt{2}(\vert{\leftrightarrow}\rangle_{\bf s}                    +
\vert{\nearrow\!\!\!\!\!\!\swarrow}\rangle_{\bf r})    \nonumber   \\
\vert D \rangle_{i}                                          =
1/\sqrt{2}(\vert{\leftrightarrow}\rangle_{\bf s}                    +
\vert{\nwarrow\!\!\!\!\!\!\searrow}\rangle_{\bf r})         \nonumber
\end{eqnarray}Note that Alice does not change their shared secrets,
she will use the {\em same} two shared sequences to transmit the
bit values to either Bob or Sonu.
Note that, the positions of the states in the two
operating sequences are not changed only the preparation of the states
are changed.
Due to  this action, both of them are
in similar position. Now if Alice randomly selects
(randomness is meant for key-splitting)
paths to transmit bit values
both of them will get  50\% bits.  So
they  have  to co-operate to access the full key. 

\paragraph*{}Splitting
the state vector into many paths and making  every  path  as  bit-carrying  
path  at  random (randomness is meant for key splitting)
  the  protocol  can  be  extended  to
distribute the key among many users. As for example, 
input  state ($0_{\circ}$ photon) 
can  be  split up into  three  parts  {\bf  r,  s} and {\bf t} 
 (1:1:1) in four following ways, simply using  triple-slit and single 
polarization rotator.
 \begin{eqnarray} \vert A \rangle_{i}  =
1/\sqrt{3}(\vert{\leftrightarrow}\rangle_{\bf r}
                    +     \vert{\leftrightarrow}\rangle_{\bf s}     +
\vert{\leftrightarrow}\rangle_{\bf t})          \nonumber          \\
\vert B \rangle_{i}                                          =
1/\sqrt{3}(\vert{\leftrightarrow}\rangle_{\bf r}                    +
\vert{\leftrightarrow}\rangle_{\bf s}                               +
\vert{\updownarrow}\rangle_{\bf t})           \nonumber            \\
\vert C \rangle_{i}                                          =
1/\sqrt{3}(\vert{\leftrightarrow}\rangle_{\bf r}                    +
\vert{\leftrightarrow}\rangle_{\bf s}                               +
\vert{\nearrow\!\!\!\!\!\!\swarrow}\rangle_{\bf t})   \nonumber    \\
\vert D \rangle_{i}                                          =
1/\sqrt{3}(\vert{\leftrightarrow}\rangle_{\bf r}                    +
\vert{\leftrightarrow}\rangle_{\bf s}                               +
\vert{\nwarrow\!\!\!\!\!\!\searrow}\rangle_{\bf t})
     \nonumber    \end{eqnarray} These states can be used to prepare two 
operating sequences. The   density   matrix of the 
 sequence of states $\vert A \rangle_{i} $ and $\vert B \rangle_{i}$ (1:1) and  
the sequence of states
$\vert C \rangle_{i} $ and $\vert D \rangle_{i} $
(1:1)  in   the
representation {\em R} corresponding to the following sequence of
base states-\\
$ \left\{\vert{\leftrightarrow}\rangle_{\bf r}                     ,\,
\vert{\updownarrow}\rangle_{\bf r}                                ,\,
\vert{\leftrightarrow}\rangle_{\bf s}
,\,\vert{\updownarrow}\rangle_{\bf s}                             ,\,
\vert{\leftrightarrow}\rangle_{\bf t}
,\,\vert{\updownarrow}\rangle_{\bf t} \right\}$ - is,  \begin{eqnarray}
\rho  =1/6\left(\begin{array}{cccccc} 2 & 0 & 0 & 0 & 0 & 0 \\0 &
0 & 0 & 0 & 0 & 0 \\0 & 0 & 2 & 0 & 0 & 0 \\0 & 0 & 0 & 0 & 0 & 0
\\0 & 0 &  0  &  0  &  1  &  0  \\0  &  0  &  0  &  0  &  0  &  1
\end{array}\right)\nonumber\,\, \end{eqnarray} Three receivers are at 
the end of three paths. Here bit-carrying
path  is {\bf t}, so only receiver on path {\bf t} will get the bit.
If Alice randomly changes  the bit-carrying path giving equal importance
to each path, then each of
the three receivers will get
 33.33\%  bits of the key.  Of course,
 who will get the bit it is his/her duty to pursue QA.
Here op-operation is required only to access the full key.
This is accomplished by the same apparatus.
Therefore, the protocol will be more powerful  than other
key splitting protocols
which use separate
apparatus, since any denial of receiving bit values can be legally challenged by
sender and other receiver(s). Clearly this advantage arises from superposition states.
In  each bit level,
co-operation can be guaranteed in our entanglement-based scheme [14].

\paragraph*{}
In all the above protocols, states are mutually non orthogonal and 
density matrices of the 
two sequences are same. Mutual nonorthogonality and equivalence of density 
matrices are the most powerful combinations for this 
alternative QKD on disentangled state. 
Suppose the two sequences, having same density matrices,
($\rho_{0}=\rho_{1}$)
are prepared by two different pair of orthogonal
states (say, one sequence is made by $0^{\circ}$ and $90^{\circ}$ and the
other is made by
$45^{\circ}$ and $135^{\circ}$ 
polarized states). This is another interesting quantum channel
however, but so far security
is concerned this will be a weak quantum channel.
Intercepting
a single sequence, Eve
 will  not get the bit value but  can  evade detection
 if she fortunately
chooses the correct orthogonal basis. 
On the other hand, in a protocol where
two sequences of two same or different pairs of non orthogonal states,
having
unequal density matrices,($\rho_{0}\neq\rho_{1}$) are used,
  Eve can get the bit value
from the single sequence but cannot evade detection
( for the above protocols 
$\rho_{0} \neq \rho_{1}$ if coherent states are used). 
If two criteria are imposed i.e. density matrices are same and
 states are non orthogonal for both sequences, then Eve will not get the
bit value still she will introduce error. As if bank robber is
caught as soon as he/she proceeds towards the bank.
The system is extremely sensitive.
In contrast, in the existing QKD protocols eavesdropping is  detected 
only after 
leakage of  some of the bit values. \\

On the basis of
indistinguishability of differently prepared density matrix,
 once Park claimed [15] to have refuted the
legitimacy of  individual quantum state description. 
Here we saw that {\em same} density matrix can be quantum mechanically distinguished
(in his book, D'espagnat [16] presented a classical separation method).
But we would not like to make any counter claim, rather we leave  the issue of
quantum state description to be freshly reviewed in the light of this information
proccessing.\\

\noindent
{\bf Acknowledgement}: I thank C. H. Bennett, L. Goldenberg, P. Kwiat, 
H.-K. Lo,
D. Mayers, T. Mor, A. Peres, S. Popescu, P. W. Shor,
L. Vaidman, S. Wiesner and W. K. Wootters
 for help and comments at different stages of this work 
when it was partly obscure and partly misleaded.\\

\begin{table}\bf Table 1. Joint probabilities when $DA$
at $(0^{\circ}_{r}
 :   0^{\circ}_{s})$\begin{center}  \begin{tabular}{|c|cccc|}\hline
States              &              ~~~~~$P_{(\surd_{r}:\times_{\bf s})}$
&~~~~~~~$P_{(\times_{\bf r}:\surd_{\bf s})}$
~~~~~~~&$P_{(\times_{\bf r}:\times_{\bf s})}$       &       \\        \hline
$\vert A \rangle_{i}$ = $1/\sqrt{2}(\vert{\leftrightarrow}\rangle_{\bf r}                   +
\vert{\leftrightarrow}\rangle_{s})$ & $1/4$ &  $1/4^{*}$  &  $0$  &\\
$\vert B \rangle_{i}$ = $1/\sqrt{2}(\vert{\leftrightarrow}\rangle_{r}                   +
\vert{\updownarrow}\rangle_{\bf s})$ & $1/4$  &  $0$  &  $1/4$  &  \\
$\vert C \rangle_{i}$ = $1/\sqrt{2}(\vert{\leftrightarrow}\rangle_{\bf r}                   +
\vert{\nearrow\!\!\!\!\!\!\swarrow}\rangle_{\bf s})$ &  $1/4$  &$1/8$
&$1/8$   &   \\  
$\vert D \rangle_{i}$ = $1/\sqrt{2}(\vert{\leftrightarrow}\rangle_{\bf r}  +
\vert{\nwarrow\!\!\!\!\!\!\searrow}\rangle_{\bf s})$ &  $1/4$  &$1/8$
&$1/8$ & \\ \hline \end{tabular}\end{center}
 \small  * Only this  result ($\beta$)  provides
conclusive $WPWS$ information. \end{table} 

\begin{table}\bf Table  2.  Joint    probabilities    when    $DA$    at    $(
0^{\circ}_{r}:45^{\circ}_{s})$\begin{center}
\begin{tabular}{|c|cccc|}\hline             States            &
~~~~~$P_{(\surd_{\bf r}:\times_{\bf s})}$ &
~~~~~~~$P_{(\times_{\bf r}:\surd_{\bf s})}$
~~~~~~~&$P_{(\times_{\bf r}:\times_{\bf s})}$ & \\
     \hline   
$\vert A \rangle_{i}$ =  $1/\sqrt{2}(\vert{\leftrightarrow}\rangle_{\bf r}    +
\vert{\leftrightarrow}\rangle_{\bf s})$ & $1/4$ & $1/8$ &  $1/8$  &\\
$\vert B \rangle_{i}$ = $1/\sqrt{2}(\vert{\leftrightarrow}\rangle_{\bf r}                   +
\vert{\updownarrow}\rangle_{\bf s})$ & $1/4$  &  $1/8$  &  $1/8$&  \\
$\vert C \rangle_{i}$ = $1/\sqrt{2}(\vert{\leftrightarrow}\rangle_{\bf r}                   +
\vert{\nearrow\!\!\!\!\!\!\swarrow}\rangle_{\bf s})$ & $1/4$ &  $1/4^{*}$
&   $0$   &\\  
$\vert D \rangle_{i}$ =  $1/\sqrt{2}(\vert{\leftrightarrow}\rangle_{\bf r}   +
\vert{\nwarrow\!\!\!\!\!\!\searrow}\rangle_{\bf s})$ & $1/4$&  $0$  &
$1/4$ & \\ \hline \end{tabular}\end{center}
 \small * Only this  result ($\beta$) provides
conclusive $WPWS$ information.

\end{table}

\end{document}